# RF Energy Absorption in Human Bodies due to Wearable Antennas in the 2.4 GHz Frequency Band

Marta Fernandez[1], Hugo G. Espinosa[2], David Guerra[1], Iván Peña[1], David V. Thiel[2] and Amaia Arrinda[1]

[1]Department of Communications Engineering, University of the Basque Country (UPV/EHU), Bilbao, Spain

[2]School of Engineering, Griffith University, Nathan Campus, Brisbane, Australia

Corresponding author:

Dr. Marta Fernandez

Department of Communications Engineering,

University of the Basque Country (UPV/EHU),

Alda. Urquijo s/n, Bilbao 48013, Spain

Email: martafernandez010@gmail.com

Phone: +34 946017381

Grant sponsors: Basque Government (IT-683-13), University of the Basque Country UPV/EHU under Grant Dokberri 2018-II (DOCREC18/36), and Erasmus Mundus PhD and Postdoctoral exchange program under the PANTHER project.

**Abstract:** Human exposure to electromagnetic fields produced by two wearable antennas operating in the 2.4 GHz frequency band was assessed by means of computational tools. Both antennas were designed to be attached to the body skin, but they were intended for different applications. The first antenna was designed to be used in off-body applications, this is, to communicate with a device placed outside the body; while the second antenna model was optimized to communicate with a device located inside the body, and thus, to establish in-body communications. The power absorption in human tissues was determined at several locations of adult male and female body models. The maximum SAR value obtained with the antenna designed for off-body applications was found on the torso of the woman model and was equal to 0.037 W/kg at 2.45 GHz. SAR levels increased significantly for the antenna transmitting inside the body. In this case, SAR values ranged between 0.23 W/kg and 0.45 W/kg at the same body location. The power absorbed in different body tissues and the total power absorbed in the body were also calculated, being this maximum total power absorbed equal to 5.2 mW for an antenna input power equal to 10 mW.

**Introduction**

Wearable antennas and implantable devices are increasing in demand because of their utility and application in various fields, such as medicine, sports or entertainment [Einicke et al., 2017; Kiourti et al., 2014; Nemati et al., 2012]. The different devices of a Wireless Body Area Network (WBAN) establish on-body communications when the communicating devices are on the body; in-body communications when at least one of the devices is inside the body; and off-body communications when a device placed on the body communicates with an external source. One of the main challenges when designing a wearable antenna is to keep the efficiency of the antenna when it is placed on the body while maintaining low power absorption in the human tissues. The impact of the body on the antenna performance has been analyzed for different antennas designed for off-body and on-body applications, such as in the work developed by Jiang and Werner [2015]. Some of the drawbacks of the different wearable antenna designs are described in [Varnoosfaderani et al., 2015a]. The development of antennas for in-body communications, required for biomedical purposes, is another challenge, since the antenna has to match the body impedance. The capsule endoscopy is an example of application that requires an ingestible pill and a receiver or a transceiver out of the body [Kim et al., 2012].

The power absorption in human tissues is often investigated in order to check compliance with the standards and regulations [ICNIRP, 1998; IEEE, 2005], especially when the radiofrequency (RF) source is close to the human body. This is the case of user devices, such as mobile phones, laptops or tablets, whose radiation can be evaluated by means of certain standardized laboratory measurements in order to certify the device [Davis and Balzano, 2009; Deltour et al., 2011]. A more detailed research is conducted using computational tools or measurements with the aim of investigating the exposure due to personal devices in different situations, such as inside vehicles [Leung et al., 2012] or considering different postures of a person [Krayni et al., 2017]. Antennas located inside the body, on the body or close to it are emerging technologies that also require electromagnetic (EM) compliance testing. For instance, in [Bercich et al., 2013] the authors investigated the possibility of remotely powering devices implanted in the body tissues without exceeding the safety limitations.

Numerical methods are widely used in EMF exposure assessment and in dosimetry studies and they can be applied to a wide range of analysis. For example, numerical calculations were used for assessing the absorption and distribution of EMF levels in different body tissues [Alekseev and Ziskin, 2009], to study the influence of the permittivity on specific absorption rate (SAR) calculations [Hurt et al., 2000] and for investigating the safety distance limits for a human near a mobile phone base station [Cooper et al., 2002].

Recently, the authors developed a wearable slot antenna for off-body communications working at 2.45 GHz [Fernandez et al., 2017], showing a good agreement between antenna performance and body absorption. Applications of this antenna model can be related to sport training or entertainment, in which an on-body sensor communicates with the exterior. As specified by Sabti et al. [2015], body sensors allow professionals to effectively measure the athlete training sessions and perform early diagnosis and treatment. For example, in [Marin-Perianu et al., 2013] wearable sensors together with a wireless device located out of the body were used to allow the cyclists to get immediate feedback about their pedaling movement. Moreover, this antenna intended for off-body communications can also be used for remote healthcare monitoring. For example, Abidoye et al. [2011] proposed a system architecture for remote healthcare monitoring, where one of the sensors of the WBAN communicates with a personal server using ZigBee technology. The advantage of the antenna presented in this paper lies in the improvement made in order to minimize the power radiated towards the body.

Furthermore, the authors redesigned this antenna for inward radiation applications [Fernandez et al., 2018]. In this case, the antenna was placed on the body skin for communication with a radio capsule inside the body. The capsule endoscopy permits to observe the complete length of the gastro-intestinal track [Basar at al., 2012]. However, one of the main issues that must be overcome in order to have wider clinical applications of wireless capsule endoscopy, is the localization of the capsule without the need to employ cables in the transceiver system [Basar et al., 2013]. Little research has been done regarding the wireless antenna placed on the body. The study presented in this paper gives solution to this problem.

Human exposure to electromagnetic fields from two wearable slot antennas is assessed by means of computational tools. The study was developed when the antennas were placed on different locations of male and female body models, both antennas were operating in the near field region. Results were given in terms of SAR levels, power absorbed in different body tissues and Whole-body SAR ($SAR_{WB}$) values.

**Methodology**

*Wearable antenna design*

Two wearable antennas working in the 2.4 GHz frequency band were used for the analysis of power absorption in the human body. Both antennas were previously designed using computational tools and fabricated with the aim of evaluating usefulness for WBANs. Simulations were carried out to verify antenna properties and its performance when it was placed on the body, since the interaction between the body and the antenna can cause a significant decrease in the antenna efficiency. After fabricating the antennas, they were tested on different bodies, measuring return losses, with the objective of validating the simulations. The type of antennas used in this work are composed by a resonant cavity with a slot. They have been used in different sports and medical applications [Varnoosfaderani et al. 2015b; Xia et al., 2009; Haga et. Al, 2009]. The use of this type of designs proved to minimize the negative effects of the body on the antenna performance [Varnoosfaderani et al., 2015b]. Furthermore, the antenna used in this study improves the performance of previous antennas of this type. As can be seen in [Fernandez et. Al 2018], our design shows an improvement of 20% in the radiation efficiency with respect to a different resonant cavity with a slot presented in [Varnoosfaderani et al., 2015b].

The first antenna model (Fig. 1a) was presented previously by the authors [Fernandez et al., 2017] and it was designed for using it in off-body applications, since it can be placed on the skin and communicate with external sources. This antenna was optimized to minimize the power radiated towards the body. It was fabricated for improving its efficiency with respect to previous wearable slot antennas, while decreasing the power absorbed in the human body. This antenna consisted of a conductive box with a slot on the lid facing out of the body. A rectangular monopole made of brass (thickness 0.1 mm and conductivity $\sigma=1.57\times10^7$ S/m) was contained inside the box to feed the antenna. The box was made of biodegradable PLA (Polylactic acid) material (PLA filament 1.75 mm, Bq, Navarra, Spain) using 3D printing technology (Bq Witbox 3D printer, Bq, Navarra, Spain). The relative permittivity $\varepsilon_r$ and the dielectric loss tangent $\tan\delta$ of the PLA were $\varepsilon_r=4$ and $\tan\delta=0.02$, respectively. Once the box was printed, the internal walls were covered with conductive silver paste (Archeson Electrodag 479SS, Henkel, Sydney, Australia) except the slot. The conductivity of the silver paste was equal to $4.3\times10^6$ S/m. The box and slot dimensions were $56\times33\times11$ mm and $47\times9$ mm, respectively, and the thickness of the box walls was 1.5 mm. As shown in Figure 1a, the slot was folded onto the side walls of the box, since its length was bigger than the width of the box. Finally, a SMA connector (SMA 8500-0000, RF Shop, Lonsdale) was attached to perform the experimental measurements.

The second antenna followed the design of the first one (see Fig. 1b), but in this case, the slot was pressed against the body skin to be used for in-body communications. Having the slot touching the body skin makes this antenna suitable for communicating with devices inside the body, so hereinafter it will be referred to as antenna design for in-body communications. In order to have this second design working in the 2.4 GHz frequency band, the box and slot dimensions, as well as the feed were optimized to match the body impedance. The dimensions of this second box were $33\times33\times11$ mm. In this case, the slot was 28 mm long and 7 mm wide. The antenna was also fabricated using 3D printing technology and the same materials used for the previous design were also used here. More details regarding this second antenna can be found in [Fernandez et al., 2018].

The two antennas were designed using the commercially available EM simulation software CST [2016] (CST Computer Simulation Technology, Munich, Germany) and a voxel body model (Gustav) included in the library of software. Both antennas were fabricated and tested on different bodies for people with different body mass index. This ensured that the antennas were working properly in a real environment. One important parameter that was analyzed by means of experimental measurements was the return loss (S11) of each antenna. The return loss, as well as the power absorbed in the body, is influenced by the dielectric properties of the body. Results from [Fernandez et al., 2017] showed that the resonant frequency of the antenna designed for off-body applications ranged between 2.262 GHz and 2.320 GHz in computational simulations when it was located at the different positions of the body. In experimental measurements, the mean resonant frequency ranged between 2.384 GHz and 2.438 GHz, and the standard deviation ranged between $6.67 \times 10^{-3}$ GHz and $3.3810^{-2}$ GHz. This standard deviation was due to the different individuals that participated in the experiment measurements. With regard to the antenna designed for in-body applications [Fernandez et al., 2018], the resonant frequency ranged between 2.41 GHz and 2.65 GHz in simulations and between 2.27 GHz and 2.72 GHz in measurements. Differences between simulations and measurements can also be due to variations in material properties, like those due to silver paste thickness and curing process of the conductive silver paste. Overall, it was shown that the different bodies of the participants had little effect on the performance of the antennas.

Regarding exposure assessment, only the off-body antenna design was tested [Fernandez et al., 2017]. A more detailed analysis of the power absorbed due to this antenna, as well as the full evaluation of the second antenna design are presented in this paper.

In addition to the extension of exposure evaluation to more locations in the body, the main features of the present work that improve the previous study are the inclusion of two body models, a male and a female, with higher resolution. Moreover, the exposure evaluation is performed at three different frequencies (2 GHz, 2.45 GHz, and 3 GHz), because the interaction of the antenna with the human body causes a shift in the resonant frequency, so the assessment at close frequencies is of interest. Another novelty of this work is the analysis of the difference in power absorption between an off-body antenna design and its equivalent for in-body radiation, together with the power absorbed in different tissues due to these types of antennas.

*Methods for evaluating exposure of wearable antennas*

The body models used in this study are the AustinMan and AustinWoman high fidelity anatomical voxel models, both of which are open source models constructed from the Visible Human Project (VHP) datasets [Massey and Yilmaz, 2016]. The resolution of the voxels is $2\times2\times2$ mm$^3$. The same computational tool than the one used for designing the antennas was employed for calculating the power absorption in body tissues, using the finite-difference time-domain (FDTD) method. For the simulations, hexahedral mesh type was used. The mesh setting was established in the range $\lambda_0/15$ - $\lambda_0/20$, where $\lambda_0$ is the free space wavelength.

Each antenna was placed at 8 different positions of the human body (Fig. 2). For the off-body antenna design the box was in contact with the skin, being the face of the slot facing out of the body. Figure 3 shows this antenna when it is placed on the leg. The antenna designed for in-body radiation was attached to human body with the side of the slot touching the skin. A simulation per antenna and body location was performed, repeating the process for the two body models. The dielectric properties of body tissues were recalculated using the 4-Cole-Cole formulation in the frequency band of interest [IFAC, 1997]. Two different computational analyses were performed: the SAR, which is a measure of the power absorbed per unit of mass, and the total power absorbed in the human body. The SAR was averaged over 10 g of contiguous tissue and the maximum value was reported in each case.

**Results of personal exposure from wearable antennas**

*SAR averaged over 10 g of mass*

The SAR values obtained when placing the antennas on the different parts of the body models are presented in Figure 4a for the antenna for off-body communications, and in Figure 4b for the antenna intended for in-body applications. The maximum SAR values obtained at 2 GHz, 2.45 GHz, and 3 GHz are given for each body model, man (M) or woman (W), and for each body location. The input power was 10 mW. As shown in both figures, when using the antenna designed for off-body communications, the maximum SAR value was equal to 0.0369 W/kg, this was obtained when placing the antenna on one side of the torso of the woman model. The minimum SAR level due to this antenna at 2.45 GHz was 0.0113 W/kg, obtained when it was placed on the woman's leg (Thigh 2). When using the antenna intended for in-body communications the SAR levels increased significantly, taking values between 0.2278 and 0.4479 W/kg at 2.45 GHz, getting this highest value also on the side of the torso of the woman body. The SAR value at 3 GHz was sometimes higher than at 2.45 GHz. In fact, the maximum SAR value at 3 GHz was 0.4686 W/kg, acquired when placing the antenna on the woman's leg. This is due to the frequency shift produced because of the different biological properties.

With regards to the exposure limits at these frequencies, the basic restrictions for the general public when the SAR is averaged over 10 g of mass are 4 W/kg for the limbs and 2 W/kg for the head and trunk, as indicated in [ICNIRP, 1998; IEEE, 2005]. Considering that the maximum SAR value was acquired on the trunk and it was equal to 0.4479 W/kg for an input power of 10 mW, the maximum permitted input power for the in-body antenna should be lower than 44.65 mW so as not to exceed the 2 W/kg indicated in the guidelines [ICNIRP, 1998]. However, for the off-body antenna, this input power can be much higher but even with 500 mW the limits are fulfilled (the input power should be lower than 542 mW).

### *Total power absorbed in the human bodies*

The total power absorbed in the body was also reported for an antenna input power of 10 mW. These 10 mW refer to the input power of the antenna, so mismatch losses are not included. Thus, the radiated power is lower than 10 mW. However, the mismatch losses are minimal at the operating frequency as can be seen in [Fernandez et al., 2017; Fernandez et al., 2018], where the values of the $S_{11}$ parameter of the antennas were presented. Regarding the antenna design made for off-body applications, the highest power absorbed at 2.45 GHz was found on the side of the torso of the woman and man voxel models, being equal to 5.21 mW and 3.85 mW, respectively. When placing the antenna at different locations in the leg, the maximum power absorbed at the same frequency was 3.84 mW on the Thigh 2 position of the man model. Regarding the arm positions, the highest value was 3.41 mW on the Arm 2 location of the woman model. In relation to the in-body antenna, the maximum power absorbed by the body at 2.45 GHz was found when placing this antenna on the Thigh 1 position of the woman, reaching a value of 9.78 mW. The highest levels obtained on the torso and arm locations were 9.62 mW (Torso 2 M) and 9.50 mW (Arm 2 W), respectively. The maximum power levels obtained on the torso, leg, and arm at 2.45 GHz, which is the frequency where the most significant results were obtained, are summarized in Table 1, detailing the specific location and the body model where these levels were obtained. The total power absorbed in the body was sometimes higher at 3 GHz than at 2.45 GHz for the in-body antenna because of the frequency shift. The highest power absorbed at 3 GHz was 9.75 mW and it occurred when the antenna was placed on Thigh 2 of the man model.

***Power absorbed in the different tissues***

The power absorbed in the different tissues was analyzed for each body position at the three frequencies. Figure 5 shows the percentage of power absorbed in each tissue due to the off-body and in-body antennas when placing them on the man (Man Off, Man IN) and woman (Woman Off, Woman IN) models. For both antennas, the highest amount of power was absorbed in the three first layers of the human models (skin, fat, and muscle). This is due to the low penetration depth at these frequencies. Overall, when the antenna was on the torso, the skin and fat absorbed more power than the muscle, being the energy absorbed by this latter layer lower than the 26% of the total power absorbed in the body. On the arm and on the lower leg, the percentage of radiation absorbed in the fat was lower, since in these parts the fat layer is thinner. In these locations, an increase in the power deposited in the muscle was determined, with values between 25% and 73%. When the antenna was placed on the limbs, the power absorbed in the tendon was also noticeable in some cases, for example, at the Arm 2 position of the woman body or on the lower leg.

When the antenna was on the side of the torso, some radiation was found on the bones, especially when using the antenna intended for off-body communications. This was due to the proximity of the arm to the antenna. Finally, some power was deposited in other body tissues, represented by 'Others' in the graphics, such as in the liver or stomach when placing the antenna on the torso, or in the nerves, blood vessels or bladder.

**Whole-Body SAR Results due to Wearable Antennas**

Regarding the simulations of the wearable antennas, the maximum power level absorbed in the body was equal to 5.21 mW in the case of the woman and 3.85 mW for the man body model when employing the off-body antenna design. When using the antenna intended for in-body communications, the maximum power absorbed in the body was 9.78 mW for the woman and 9.75 mW for the man body model. The total masses of the human models are 84.82 kg for the woman and 106.17 kg for the man [Massey and Yilmaz, 2016]. Calculating the $SAR_{WB}$ as the power absorbed divided by the total mass gives maximum $SAR_{WB}$ values of 36.26 µW/kg and 61.42 µW/kg for the off-body antenna and 91.83 µW/kg and 115.3 µW/kg for the antenna designed for in-body applications. The basic restriction for the $SAR_{WB}$ at these frequencies is 0.08 W/kg [ICNIRP, 1998], so all the values obtained in this analysis are below the limits.

**Conclusions**

The evaluation of near field exposure due to two wearable slot antennas working in the 2.4 GHz frequency band was performed. The reason for choosing this type of antennas was that a slot on a conductive box was considered appropriate for reducing the interaction between the human and the antenna. The first antenna model was optimized to minimize the power radiated towards the body, increasing in this way the antenna efficiency in comparison with other wearable antennas. These features make this antenna suitable for communicating with devices located out from the body, which is of interest in sports training, entertainment or health monitoring applications. The second antenna model was designed to be placed on the body skin so it can communicate with a transmitter inside the body. Little research has been done on the wireless transceiver located on the skin for in-body communications, so the antenna presented in this paper can be a solution for different biomedical applications, such as the capsule endoscopy.

Computational tools were used to evaluate power absorption due to the antennas in several locations of two different body models, a male and a female. Although the antennas were working at 2.45 GHz, a small frequency shift was produced from one location to another, so the evaluation of exposure was performed at three different frequencies (2 GHz, 2.45 GHz, 3 GHz). SAR levels averaged over 10 g of contiguous tissue were calculated, as well as the total power absorbed in each body model and the percentages of power level absorbed by the different tissues. Differences in the results were found due to the two types of antennas. In the case of the antenna aimed at radiating out of the body, the maximum SAR level calculated at 2.45 GHz was equal to 0.0369 W/kg, while when using the antenna designed for establishing in-body communications the maximum SAR level increased up to 0.4479 W/kg at the same frequency.

Exposure levels obtained in this study were compared to exposure limits for general public given in ICNIRP Guidelines [ICNIRP, 1998]. The results obtained fulfill these limitations in all cases. However, the antenna designed for in-body communications is targeted to medical applications, so in this case, general limits may not apply. In this regard, the ICNIRP Statement [ICNIRP, 2017] includes a review of the relevant regulations related to diagnostic devices using non-ionizing radiation, which includes implantable technology or applications for patient monitoring. This statement specifies that as diagnostic procedures can confer substantial benefits to health, risk of adverse effects in patients may in some circumstances be acceptable. Despite this, as the obtained results comply with the general public limitations, they also comply with the limits specified in the regulations related to medical devices.

**References:**

Abidoye AP, Azeez NA, Adesina AO, Agbele KK, Nyongesa HO. 2011. Using wearable sensors for remote healthcare monitoring system. J Sens Technol 1:22-28.

Alekseev SI, Ziskin MC. 2009. Millimeter-Wave absorption by cutaneous blood vessels: a computational study. IEEE Trans Biomed Eng 56:2380-2388.

Basar MR, Malek F, Juni KM, Idris MS, Iskandar M, Saleh M. 2012. Ingestible wireless capsule technology: A review of development and future indications. Int J Antennas Propag 2012:1-14.

Basar MR, Malek F, Juni KM, Saleh MIM, Idris MS, Mohamed L, Saudin N, Affendi NAM, Ali A. 2013. The use of a human body model to determine the variation of path losses in the human body channel in wireless capsule endoscopy. Prog Electromagn Res 133:495-513.

Bercich RA, Duffy DR, Irazoqui PP. 2013. Far-Field RF powering of implantable devices: safety considerations. IEEE Trans Biomed Eng 60:2107-2112.

Cooper J, Marx B, Buhl J, Hombach V. 2002. Determination of safety distance limits for a human near a cellular base station antenna, adopting the IEEE standard or ICNIRP guidelines. Bioelectromagnetics 23:429-443.

CST Microwave Studio. 2016. Computer Simulation Technology. Munich, Germany.

Davis CC, Balzano Q. 2009. The international intercomparison of SAR measurements on cellular telephones. IEEE Trans Electromagn Compat 51:210-216.

Deltour I, Wiart J, Taki M, Wake K, Varsier N, Mann S, Schüz J, Cardis E. 2011. Analysis of three-dimensional SAR distributions emitted by mobile phones in an epidemiological perspective. Bioelectromagnetics 32:634-643.

Einicke GA, Sabti HA, Thiel DV, Fernandez M. 2017. Maximum-entropy-rate selection of features for classifying changes in knee and ankle dynamics during running. IEEE J Biomed Health Inform 22:1097-1103.

Fernandez M, Espinosa HG, Thiel DV, Arrinda A. 2017. Wearable slot antenna at 2.45 GHz for off-body radiation: Analysis of efficiency, frequency shift, and body absorption. Bioelectromagnetics 39:25-34.

Fernandez M, Thiel DV, Arrinda A, Espinosa H. 2018. An Inward Directed Antenna for Gastro-intestinal Radio Pill Tracking at 2.45GHz. Microw Opt Technol 60:1644-1649.

Haga N, Saito K, Takahashi M, Ito K. 2009 Characteristics of Cavity Slot Antenna for Body-Area Networks. IEEE Trans Antennas Propag 57(4):837-843.

Hurt WD, Ziriax JM, Mason PA. 2000. Variability in EMF permittivity values: implications for SAR calculations. IEEE Trans Biomed Eng 47:396-401.

ICNIRP, International Commission on Non-Ionizing Radiation Protection. 1998. Guidelines for limiting exposure to time-varying electric, magnetic, and electromagnetic fields (up to 300 GHz). Health Phys 74:494–522.

IEEE, Institute of Electrical and Electronics Engineers. 2005. IEEE Standard for safety levels with respect to human exposure to radiofrequency electromagnetic Fields, 3 kHz to 300 GHz. IEEE Standard C95.1.

IFAC Institute for Applied Physics, Italian National Research Council. 1997. Calculation of the dielectric properties of body tissues. Available from http://niremf.ifac.cnr.it/tissprop/ [Last accessed March 2018].

James DA, Leadbetter RI, Neeli MR, Burkett BJ, Thiel DV, Lee JB. 2013. An integrated swimming monitoring system for the biomechanical analysis of swimming strokes. Sports Technol 4:141-150.

Jiang ZH, Werner DH. 2015. A compact, wideband circularly polarized co-designed filtering antenna and its application for wearable devices with low SAR. IEEE Trans Antennas Propag 63:3808–3818.

Kim K, Yun S, Lee S, Nam S, Yoon YJ, Cheon C. 2012. A Design of a High-Speed and High-Efficiency Capsule Endoscopy System. IEEE Trans Biomed Eng 59:1005-1011.

Kiourti A, Psathas KA, Nikita KS. 2014. Implantable and ingestible medical devices with wireless telemetry functionalities: A review of current status and challenges. Bioelectromagnetics 35:1-15.

Krayni A, Hadjem A, Vermeeren G, Sibille A, Roblin C, Joseph W, Martens L, Wiart J. 2017. Modeling and characterization of the uplink and downlink exposure in wireless networks. Int J Antennas Propag, 2017:1-15.

Leung S, Diao Y, Chan K, Siu Y, Wu Y. 2012. Specific absorption rate evaluation for passengers using wireless communication devices inside vehicles with different handedness, passenger counts, and seating locations. IEEE Trans Biomed Eng 59:2905-2912.

Marin-Perianu R, Marin-Perianu M, Havinga P, Taylor S, Begg R, Palaniswami M. 2013. A performance analysis of a wireless body-area network monitoring system for professional cycling. Pers Ubiquitous Comput 17(1):197-209.

Massey JM, Yilmaz AE. 2016. AustinMan and AustinWoman: High-fidelity, anatomical voxel models developed from the VHP color images. Proc Int Conf IEEE Eng Med Biol Soc.

Nemati E, Deen MJ, Mondal T. 2012. A wireless wearable ECG sensor for long-term applications. IEEE Commun Mag 50(1):36–43.

Sabti HA, Thiel DV. 2015. Self-calibrating body sensor network based on periodic human movements. IEEE Sens J 15(3):1552-1558.

Varnoosfaderani MV, Thiel DV, Lu J. 2015a. External parasitic elements on clothing for improved performance of wearable antennas. IEEE Sens J 15:307-315.

Varnoosfaderani MV, Thiel DV, Lu J. 2015b. A wideband slot antenna in a box for wearable sensor nodes. IEEE Antennas Wirel Propag Lett 14:1494–1497.

Xia W, Saito K, Takahashi M, Ito K. 2009. Performances of an Implanted Cavity Slot Antenna Embedded in the Human Arm. IEEE Trans Antennas Propag 57(4):894-899.

**Figure captions**

Fig. 1. 3D models of the wearable antennas for (a) off-body radiation, (b) in-body radiation.

Fig. 2. Positions of the body models where the antennas were placed.

Fig. 3. Antenna designed for off-body applications placed on the leg of a body model.

Fig. 4. SAR values obtained for the woman (W) and man (M) voxel models at 2 GHz, 2.45 GHz and 3 GHz when using (a) the antenna for off-body applications (b) the antenna designed for in-body communications

Fig. 5. Power absorbed in each tissue for the antennas placed on different locations of the body models: (a) Torso 1 (b) Torso 2 (c) Torso 3 (d) Arm 1 (e) Arm 2 (f) Thigh 1 (g) Thigh 2 (h) Lower Leg

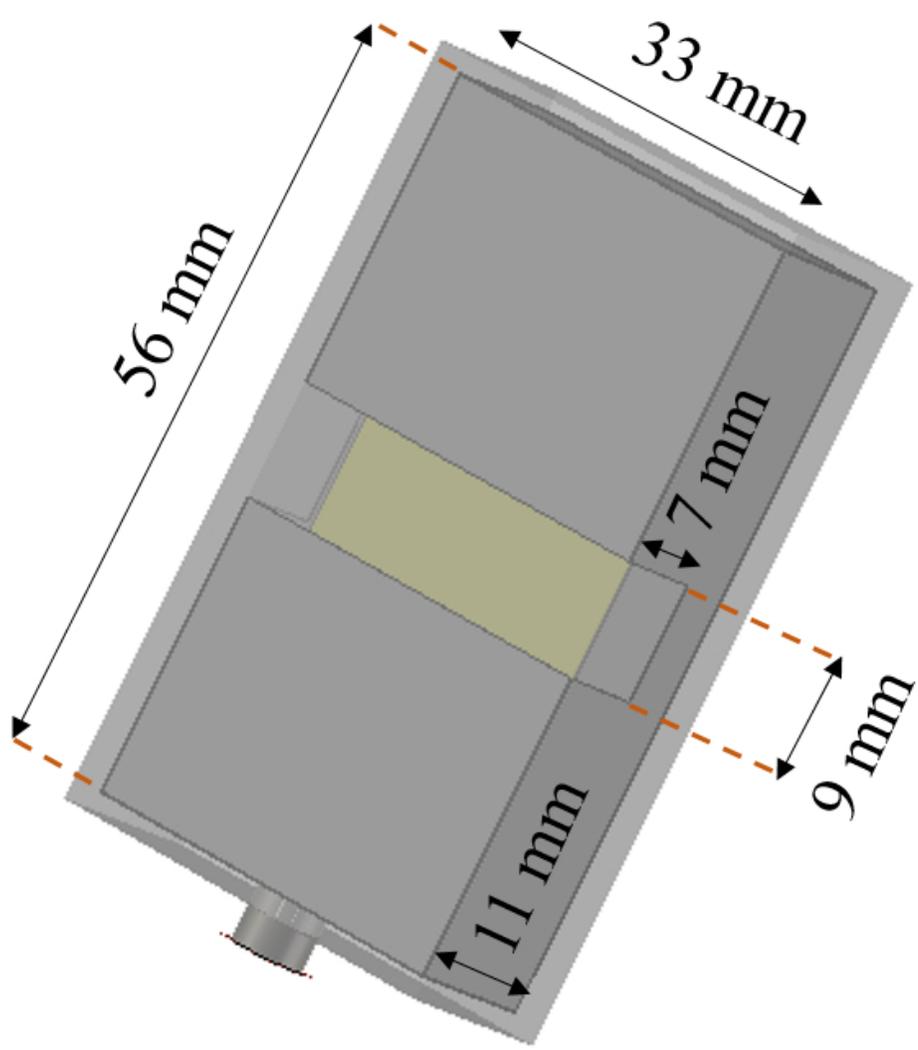

Fig. 1. 3D models of the wearable antennas for (a) off-body radiation,

51x56mm (300 x 300 DPI)

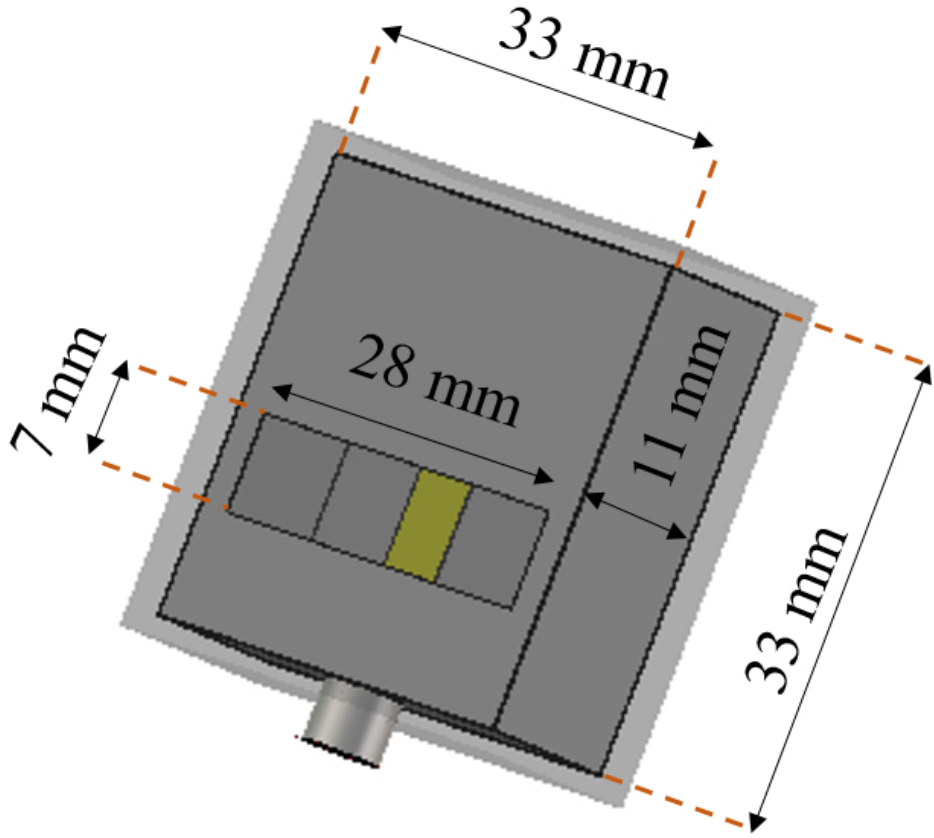

(b) in-body radiation

55x50mm (300 x 300 DPI)

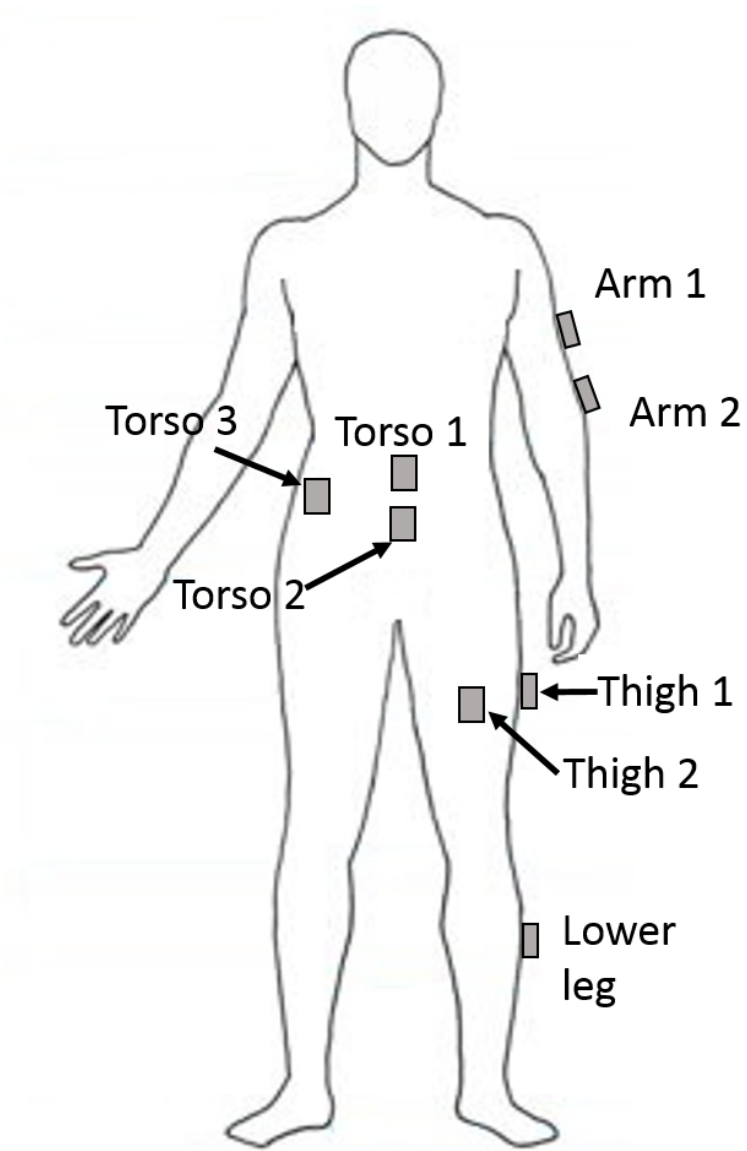

Fig. 2. Positions of the body models where the antennas were placed.

54x82mm (300 x 300 DPI)

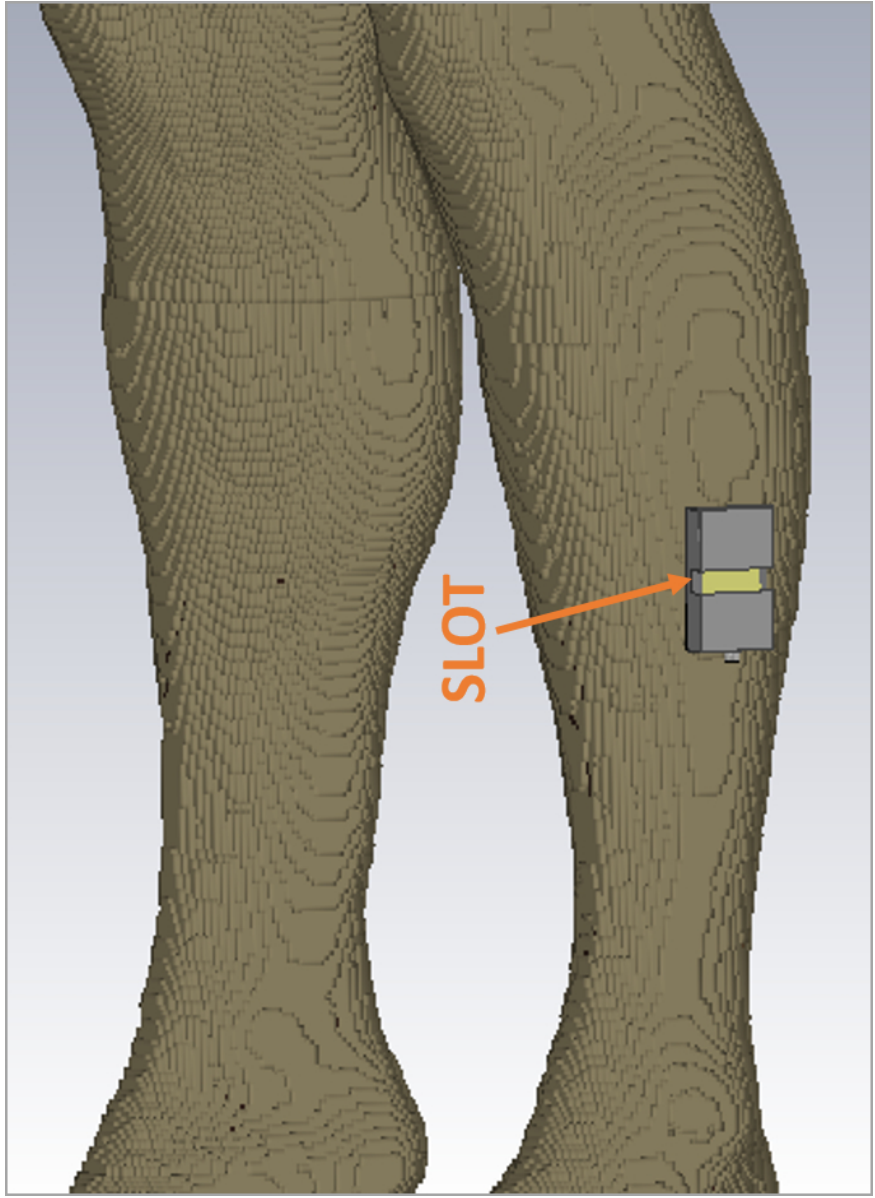

Fig. 3. Antenna designed for off-body applications placed on the leg of a body model.

57x79mm (300 x 300 DPI)

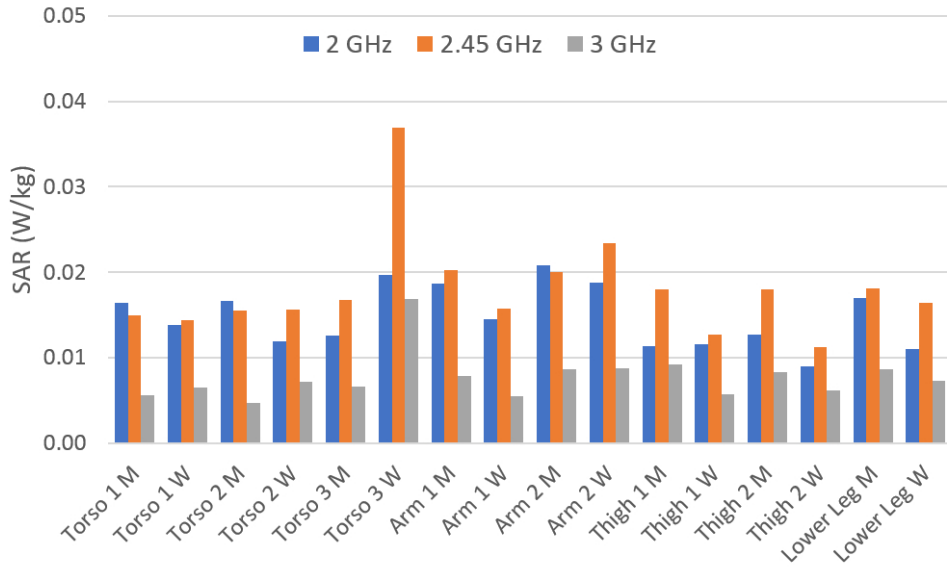

Fig. 4. SAR values obtained for the woman (W) and man (M) voxel models at 2 GHz, 2.45 GHz and 3 GHz when using (a) the antenna for off-body applications

87x51mm (300 x 300 DPI)

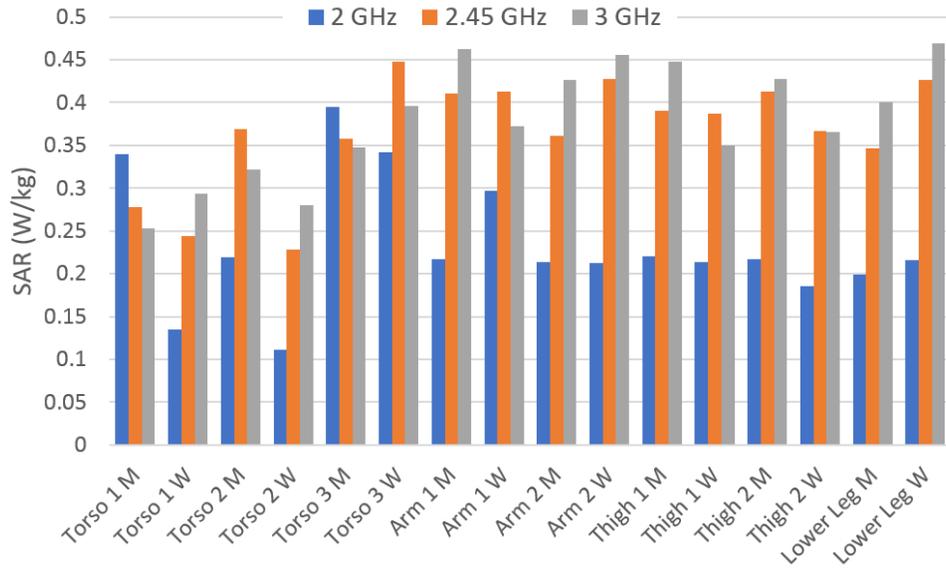

(b) the antenna designed for in-body communications

87x51mm (300 x 300 DPI)

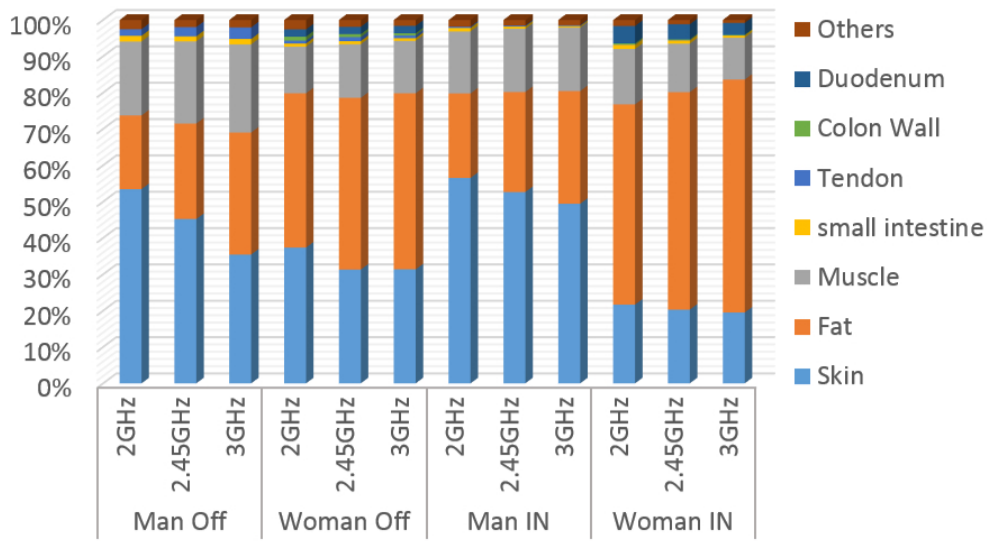

Fig. 5. Power absorbed in each tissue for the antennas placed on different locations of the body models: (a) Torso 1

75x43mm (300 x 300 DPI)

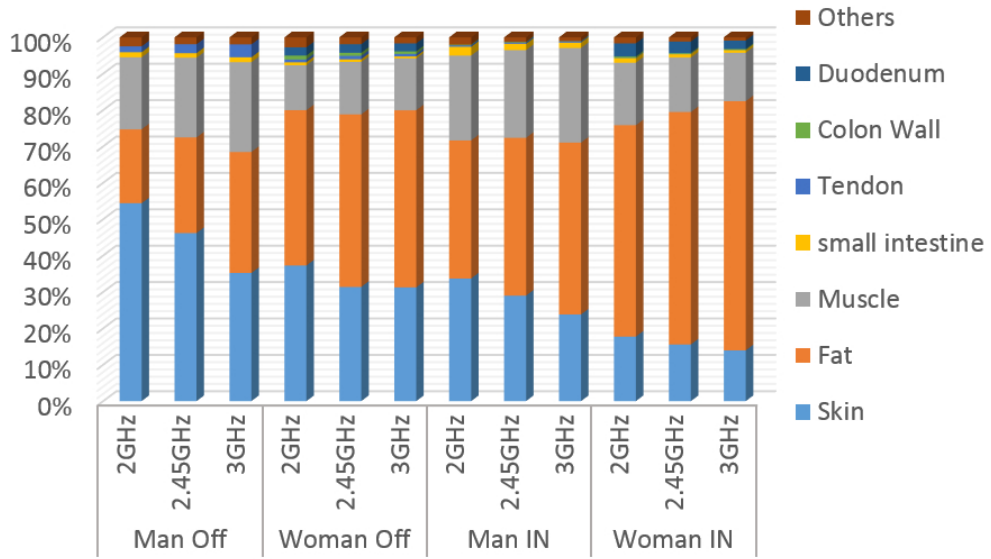

(b) Torso 2

75x44mm (300 x 300 DPI)

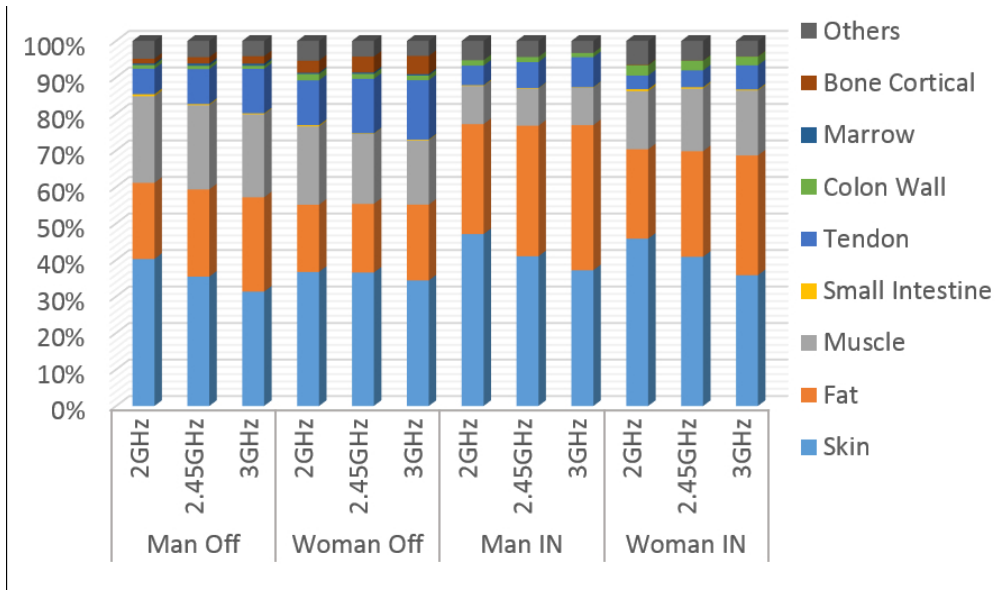

(c) Torso 3

75x44mm (300 x 300 DPI)

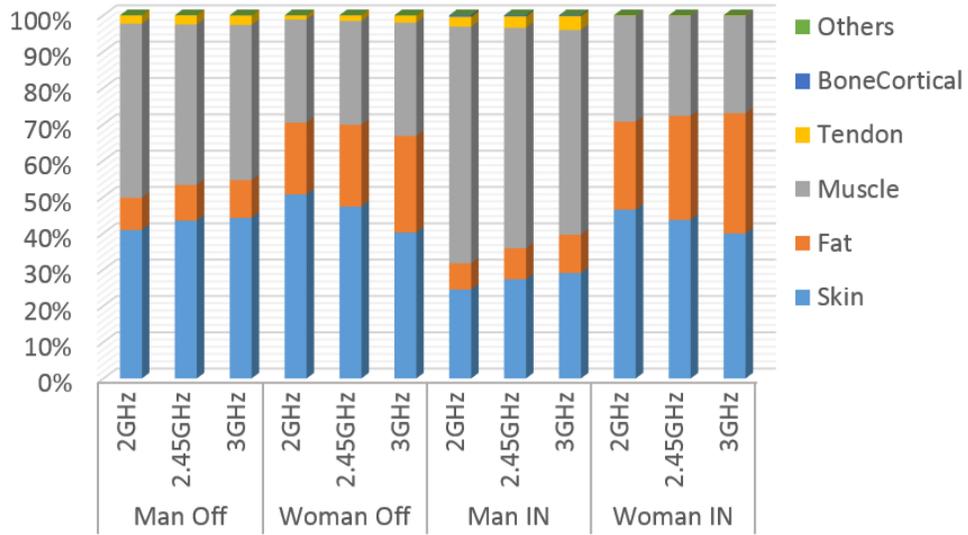

(d) Arm 1

75x43mm (300 x 300 DPI)

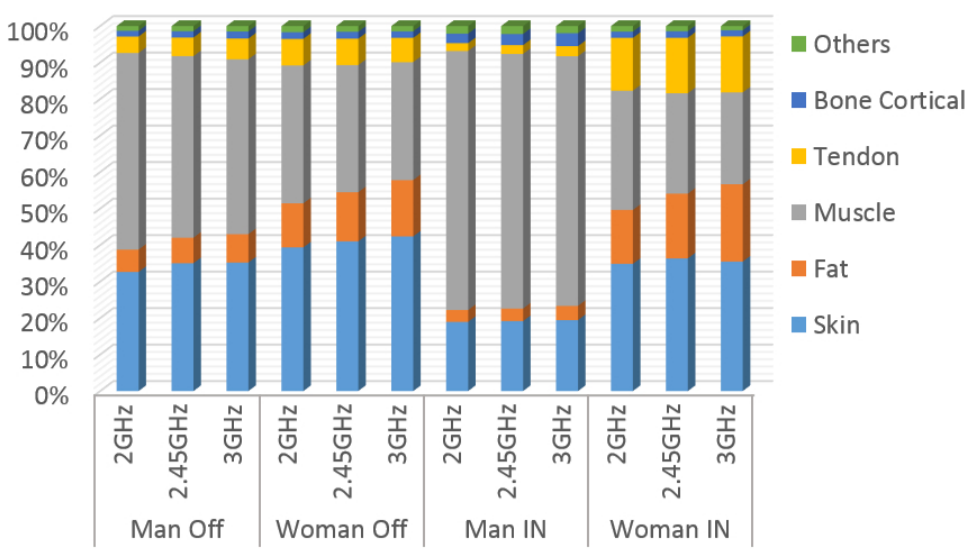

(e) Arm 2

75x44mm (300 x 300 DPI)

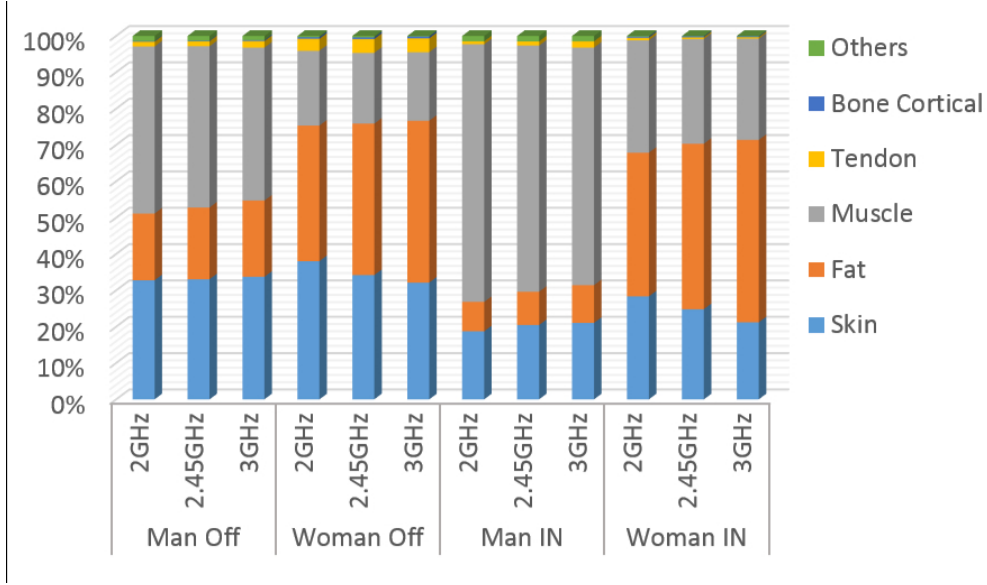

(f) Thigh 1

75x43mm (300 x 300 DPI)

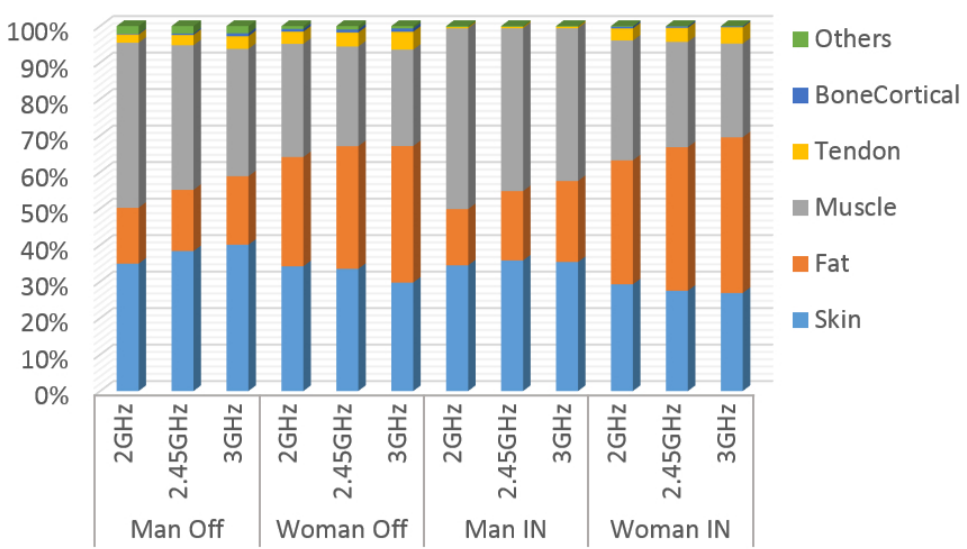

(g) Thigh 2

75x44mm (300 x 300 DPI)

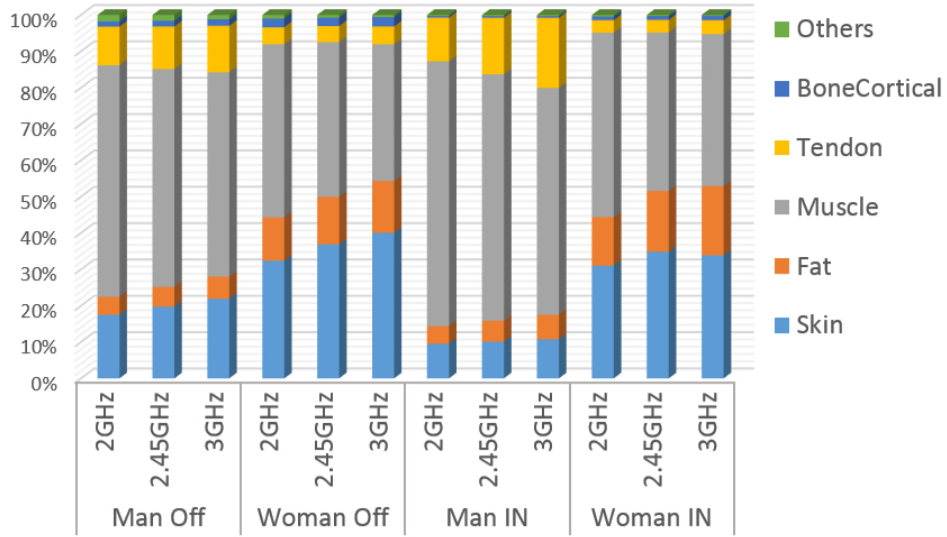

(h) Lower Leg

75x43mm (300 x 300 DPI)

Table 1. Maximum power absorbed in the different body parts at 2.45 GHz

| Body Part | Off-body | | In-body | |
|---|---|---|---|---|
| | $P_{abs}$(mW) | Location | $P_{abs}$(mW) | Location |
| Torso | 5.21 | Torso 3 Woman | 9.62 | Torso 2 Man |
| Arm | 3.41 | Arm 2 Woman | 9.5 | Arm 2 Woman |
| Leg | 3.84 | Thigh 2 Man | 9.78 | Thigh 1 Woman |